\documentclass[12pt]{article}
\usepackage{amsmath,amsfonts,amssymb,amscd}
\usepackage{color}

\numberwithin{equation}{section}
\definecolor{darkblue}{rgb}{0.02,0.548,0.908}
\definecolor{darkgreen}{rgb}{0,0.35,0}

\begin{document}

\title{Vacuum energy in asymptotically flat $2+1$ gravity}
\author{Olivera Miskovic\,$^a$\,, Rodrigo Olea\,$^b$\, and Debraj Roy\,$^a$
\vspace{1ex} \\
$^a$\textit{\small Instituto de F\'\i sica, Pontificia Universidad
Cat\'olica de Valpara\'\i so, }\\
\textit{\small Casilla 4059, Valpara\'\i so, Chile}\vspace{1ex}\\
$^{b}$\textit{{\small Departamento de Ciencias F\'{\i}sicas, Universidad
Andres Bello,}}\\
{\small \ \textit{Sazi\'{e} 2212, Piso 7,}} \textit{{\small Santiago, Chile.
\vspace{1ex}}}\\
{\small \texttt{rodrigo.olea@unab.cl, olivera.miskovic@pucv.cl,
roy.debraj@pucv.cl}}}
\maketitle

\begin{abstract}
We compute the vacuum energy of three-dimensional asymptotically flat space
based on a Chern-Simons formulation for the Poincar\'{e} group. The
equivalent action is nothing but the Einstein-Hilbert term in the bulk plus
half of the Gibbons-Hawking term at the boundary. The derivation is based on
the evaluation of the Noether charges in the vacuum. We obtain that the
vacuum energy of this space has the same value as the one of the
asymptotically flat limit of three-dimensional anti-de Sitter space.
\end{abstract}

\section{Introduction}

Asymptotically flat spacetimes are one of the most intuitive classes of
systems that exist in gravity. We expect that, for localized matter
distributions, the Einstein equations will have solutions asymptotically
matching Minkowski space, far away from the source. In four dimensions, even
outside matter distributions, the vacuum Einstein equations can accommodate
solutions with non-zero Riemann curvature, as is seen for example in the
case of the Schwarzschild black hole. The Riemann curvature there tends to
the Minkowski flat value of zero at large distances, parameterized by the
radial coordinate.

The picture in three dimensions, however, is different as gravity is
topological in nature. The Riemann tensor here has only six independent
components, and is linearly related with the Einstein tensor. The Einstein
equation necessarily gives vacuum solutions which are locally Riemann flat.
So, if the metric is to be the field that describes an isolated mass
distribution in three dimensions, the information about the mass can only
manifest as topological properties of the spacetime. Various schemes towards
this end exist. For example, in spacetimes with cosmological constant $%
\Lambda =0$, conical singularities generated by isolated mass particles have
the mass encoded in an angular deficit of the azimuthal periodicity in the
metric, which becomes less than $2\pi $ \cite{Deser:1983tn}. Also, depending
on the parameter enumerating angular deficit, one can even get solutions
which are angular excesses, though they do not represent physical solutions.
On the other hand, identification of points along the curves of a Killing
vector comprising a linear combination of Lorentz boosts and a translation
along a spatial direction has been carried out in flat space leading to
\textit{flat-space cosmologies} \cite{Cornalba:2002fi}. These topological
identifications were inspired by the ones in AdS$_{3}$ leading to the BTZ
black hole. In fact, following \cite{Barnich:2012aw}, the whole class of
solutions in $\left( 2+1\right) $-dimensional flat space is classified by
two free, dimensionless, parameters $\mu $ and $j$. With $G$ being the
three-dimensional gravitational constant, the parameter $\mu =8GM$ is
related to mass, while $j=4GJ$ is related to angular momentum.

Due to the existence of these various solutions, all of which must return to
the 3D Minkowski solution in appropriate limits of the parameters describing
the respective topological deformations corresponding to the 3D vacuum, the
role of physical properties of the vacuum itself becomes quite important. We
focus here on the vacuum energy of 3D Minkowski space. Adopting a
field-theory approach and using an off-shell equivalence between
three-dimensional Einstein-Hilbert gravity and Chern-Simons action for
Poincar\'{e} gauge group, we calculate the mass as the Noether charge for
spacetime diffeomorphisms, which is on-shell equivalent to gauge
transformations. We do this for two classes of physically admissible
solutions, the conical singularity and flat-space cosmologies.

Spacetimes whose parameters lie in a negative interval $-1<\mu =-\alpha
^{2}<0$ possess a conical defect of magnitude $2\pi (1-\left\vert \alpha
\right\vert )$. These are, in general, spacetimes of a spinning particle.
\footnote{%
In AdS$_{3}$ space, the spinning particles are nothing but the BTZ black
hole with negative mass \cite{Miskovic:2009uz}.} The static sector of a
massive point particle is given by \cite{Deser:1983tn}
\begin{equation}
ds^{2}=-dt^{2}+r^{-\beta }\left( dr^{2}+r^{2}d\theta ^{2}\right) \,,\qquad
0\leq \theta <2\pi \,.  \label{metricConical}
\end{equation}
The value of $G$ is fixed by the usual pre-factor in the Einstein equation $%
G_{\mu \nu }=8\pi G\,T_{\mu \nu }$, where the speed of light has been set to
unity. To see that this solution is locally flat, it is convenient to make a
coordinate transformation $(r,\theta )\rightarrow (\rho ,\phi )$,%
\begin{equation}
\rho =\frac{r^{\alpha }}{\alpha }\,,\qquad \phi =\alpha \,\theta \,,
\label{coordTransfconical}
\end{equation}
with $\alpha =\frac{2-\beta }{2}$, which leads to the transformed flat
metric
\begin{equation}
ds^{2}=-dt^{2}+d\rho ^{2}+\rho ^{2}d\phi ^{2},\qquad 0\leq \phi <2\pi \alpha
\,.  \label{metricConicalAdeficit}
\end{equation}%
The point to note here is the altered range of the angular coordinate $\phi $%
, modulated by the parameter $\alpha $, which describes an angular deficit
or excess, when $\alpha \neq 1$. For $\beta >2$, $\alpha $ becomes negative
and the original point $r=0$ containing the mass is mapped by (\ref%
{coordTransfconical}) to $\rho =\infty $, thus destroying the physical
picture and asymptotics. Indeed, Ashtekar et al. \cite{Ashtekar:1993ds}
noted that the points $\rho =\infty $ are at a \emph{finite} geodesic
distance away from any point in the interior. This shows the breakdown of
asymptotic flatness as the concept of being `far away' from an isolated
source. On the other hand, for $\beta <0$, or in an interval \ of parameters
$\mu =-\alpha ^{2}<-1$, the angular range exceeds $2\pi $ and instead of a
deficit, we have an \emph{excess}, describing a hyperbolic geometry similar
to lettuce leaves, which is not necessarily asymptotically flat. Thus the
range of parameters accommodating asymptotical flatness is
\begin{equation}
0<\alpha \leq 1\quad \Leftrightarrow \quad 0\leq \beta <2\,,
\label{betaRange}
\end{equation}%
where $\alpha =1$ or $\beta =0$ gives us the 3-dimensional Minkowski
spacetime. In this range, the deficit angle is related to the mass of the
particle, $m$, measured with respect to the Minkowski vacuum\footnote{%
The mass $m$ of the point particle is shifted so that the Minkowski space, $%
\mu =-1$, corresponds to $m=0$, that is, $\mu =-\left( 1-4Gm\right) ^{2}$.},
through $\beta =8Gm$. This deficit angle is always present, at any distance
from the source including at infinity, and thus the spacetime is never
asymptotically Minkowskian, unless the mass is zero. This is an important
distinction from four dimensions as, even in the leading order of an
asymptotic expansion, the spacetime is not Minkowskian and carries
information about the mass.

Investigations of spacetimes with such asymptotics give interesting results.
Ashtekar et al. \cite{Ashtekar:1993ds} considered generic asymptotically
flat spacetimes whose boundary behaviour matches that of the conical
singularity (\ref{metricConical}) and demonstrated that the bound on the
range of $\beta $ translated to the Hamiltonian being bounded \emph{both}
from above and below. Their starting point was the usual Einstein-Hilbert
action, adopting a Regge-Teitelboim \cite{Regge:1974zd} approach of adding
necessary surface terms to the Hamiltonian, which gives the conserved
quantities. The energy corresponded to a Hamiltonian that generates time
translations only for $\beta <2$, with the value of energy being positive
and lying in the range $[0,1/4G]$. The energy of the Minkowski vacuum turns
out to be zero.

Later, Marolf et al. in \cite{Marolf:2006xj} consider a finite,
differentiable action consisting of the Einstein-Hilbert term in the bulk
and the Gibbons-Hawking term at the boundary for the same asymptotics that
leads to a Hamiltonian with the same behavior of energy being bounded from
both above and below. However, the energy appears now shifted and found to
be negative, lying in the range $[-1/4G,0]$ with the energy of the Minkowski
vacuum set to $-1/4G$. Both approaches were in the metric formulation. In
contrast, Corichi et al. in \cite{Corichi:2015nea} adopted a first-order
Hamiltonian formulation and showed that the results in both references
\cite{Ashtekar:1993ds} and \cite{Marolf:2006xj} could be reproduced.

On the other hand, Barnich et al. in \cite{Barnich:2012aw} calculated the vacuum energy
as the flat limit of the cosmological AdS solutions and they found that it coincides
with the value of the AdS vacuum.

The class of conical singularities described by eq.(\ref{betaRange}) are
supplemented by two other classes of spacetimes, depending on other choices
of the parameters $\mu $ and $j$. As mentioned above, the defect $\mu
=-\alpha ^{2}<0$ corresponds to a space with an angular deficit ($\alpha
^{2}<1$) or excess ($\alpha ^{2}>1$). On the other hand, when $\mu =\alpha
^{2}>0$, these geometries can be interpreted as cosmological spacetimes. For
completeness, we note that there exist the null orbifold when $\mu =0=j$,
but we will not consider it here. Among the cases we consider, the Minkowski
space for $\left\vert \alpha \right\vert =1$ and $j=0$, is accessible as a
limit for conical singularities and angular excesses, or discretely from
flat space cosmologies. We shall discuss all asymptotics with $\mu \neq 0$.

In order to clarify the controversy in the literature about the value of the vacuum energy of Minkowski space, we adopt a field theory approach where we consider
 an action  for 3D flat gravity given by a Chern-Simons (CS) form for Poincar\'e group. The
CS action naturally comes equipped with a boundary term that is one half of
the usual Gibbons-Hawking term. This is along the line of a similar proposal for the AdS case, discussed earlier
in \cite{Mora:2004kb, Mora:2006ka, Miskovic:2006tm}. It has been recently pointed out in \cite{Detournay:2014fva} that the addition of half of the Gibbons-Hawking term on top
of the Einstein-Hilbert action has a well defined
variational principle in the asymptotically flat case.

The CS gravity action has some advantages with respect to the
Einstein-Hilbert one. For example, it is more suitable for construction of
flat space supergravity through a direct supersymmetrization of a gauge
group \cite{Nishino:1991sr,Howe:1995zm}; high-spin theory in 3D is described
by the CS action for $SL(n,\mathbb{R})\times SL(n,\mathbb{R})$ \cite%
{Afshar:2013vka}; spin-3 action in 2D can be obtained via reduction of CS
flat action with a boundary \cite{Gonzalez:2014tba}; 3D conformal gravity is
a CS theory \cite{Afshar:2013bla}, etc. On the other hand, some applications
of the CS action in 3D include a tunneling from flat space to flat space
cosmology \cite{Bagchi:2013lma} and logarithmic corrections to entropy \cite%
{Bagchi:2013qva}.

\section{Poincar\'{e} Chern-Simons gravity}

General Relativity on a $2+1$ dimensional manifold $\mathcal{M}$ can be
written as a Chern-Simons gauge theory invariant under local the Poincar\'{e}
group \cite{Witten:1988hc}
\begin{equation}
I_{CS}[A]=\frac{k}{4\pi }\int\limits_{\mathcal{M}}\left\langle A\wedge dA+%
\frac{2}{3}\,A\wedge A\wedge A\right\rangle ,  \label{actionCS}
\end{equation}%
where the constant $k$ is called the level of the theory and $\left\langle
\cdots \right\rangle $ is the trace of group generators. The gauge
connection\ 1-form $A=A_{\mu }(x)\,dx^{\mu }$ takes values in the Poincar%
\'{e} algebra $\mathfrak{iso(2,1)}$ as $A=\frac{1}{2}\,\omega
^{AB}J_{AB}+e^{A}P_{A}$. Here $\omega ^{AB}=\omega _{\mu }^{AB}(x)\,dx^{\mu
} $ and $e^{A}=e_{\mu }^{A}(x)\,dx^{\mu }$ are the gauge field 1-forms --
the spin connection and the vielbein, respectively. The Greek indices $\mu
,\nu ,\ldots =\left( t,r,\theta \right) $ label the space-time coordinates,
and the Latin ones $A,B,\ldots =0,1,2$ are the Lie algebra indices.
Furthermore, $J_{AB}$,$\ P_{A}$ are the $\mathfrak{iso(2,1)}$ generators
obeying the $2+1$ dimensional Poincar\'{e} algebra
\begin{eqnarray}
\lbrack J_{AB},J_{CD}] &=&\eta _{AD}J_{BC}-\eta _{AC}J_{BD}+\eta
_{BC}J_{AD}-\eta _{BD}J_{AC}\,,  \notag \\
\lbrack P_{A},J_{BC}] &=&\eta _{AB}P_{C}-\eta _{AC}P_{B}\,,  \notag \\
\lbrack P_{A},P_{B}] &=&0\,.
\end{eqnarray}%
We use the signature $\eta _{AB}=$diag$\left( -,+,+\right) $.\ The trace of
the above generators defines the invariant tensor of the Lie algebra and it
has the form $\left\langle J_{AB}P_{C}\right\rangle =\epsilon _{ABC}$, while
$\left\langle J_{AB}J_{CD}\right\rangle =0=\left\langle
P_{A}P_{B}\right\rangle $. With this construction for the gauge connection $%
A_{\mu }$, we see the action (\ref{actionCS}) transforming to
\begin{equation}
I_{CS}=\frac{k}{4\pi }\int\limits_{\mathcal{M}}\epsilon _{ABC}\,R^{AB}\wedge
e^{C}-\frac{k}{8\pi }\int\limits_{\partial \mathcal{M}}\epsilon
_{ABC}\,\omega ^{AB}\wedge e^{C},  \label{csactionintermediate}
\end{equation}%
where $R^{AB}=\frac{1}{2}R_{\ \ \mu \nu }^{AB}\,dx^{\mu }dx^{\nu }=d\omega
^{AB}+\omega _{\ C}^{A}\wedge \omega ^{CB}$. The first term is exactly the
Einstein-Hilbert action, once we realize that the localized gauge fields $%
\omega ^{AB}$ and $e^{A}$ are nothing but the spin connection and triad
frame fields of first-order gravity,
\begin{equation}
I_{EH}=\frac{1}{16\pi G}\int d^{3}x\,\sqrt{-g}\,R=\frac{1}{32\pi G}\int
\epsilon _{ABC}\,R^{AB}\wedge e^{C},  \label{actionEH}
\end{equation}%
identifying the level of the Chern-Simons theory with the gravitational
constant $G$ by $k=\frac{1}{4G}$.

The second term in (\ref{csactionintermediate}) is a boundary term defined
on the boundary $\partial \mathcal{M}$. We take a radial Gaussian foliation
of the spacetime in the coordinates $x^{\mu }=(x^{1},x^{i})=(r,x^{i})$, $%
i=0,2$,
\begin{equation}
ds^{2}=N^{2}(r)\,dr^{2}+h_{ij}(r,x)\,dx^{i}dx^{j}\,,  \label{radialADM}
\end{equation}%
so that the boundary is placed at constant radius $r=r_{B}$. Here, $h_{ij}$
is the induced metric on the boundary.

We work in first-order formulation where the fundamental fields are the
vielbein $e^{A}=e_{\mu }^{A}\,dx^{\mu }$ and the Lorenz connection $\omega
^{AB}=\omega _{\mu }^{AB}\,dx^{\mu }$. One possible choice of the vielbein
in the foliation (\ref{radialADM}), where the Poincar\'{e} indices split as $%
A=(1,a)$, is
\begin{eqnarray}
e^{1} &=&N\,dr\,,  \notag \\
e^{a} &=&e_{i}^{a}\,dx^{i}\,.  \label{e}
\end{eqnarray}%
The boundary vielbein $e_{i}^{a}$ is related to the induced metric by $%
h_{ij}=\eta _{ab}\,e_{i}^{a}e_{j}^{b}$, and the extrinsic curvature of the
boundary is%
\begin{equation}
K_{ij}=-\frac{1}{2N}\,\partial _{r}h_{ij}\,.  \label{Kdefn}
\end{equation}

The components of $\omega ^{AB}$ are calculated from $de^{A}+\omega
^{AB}\wedge e_{B}=0$, leading to%
\begin{eqnarray}
\omega ^{1a} &=&K^{a}\,,  \notag \\
\omega ^{ab} &=&\omega _{i}^{ab}\,dx^{i}+e^{i[a}\partial
_{r}e_{i}^{b]}\,dr\,,  \label{w}
\end{eqnarray}%
where $K^{a}=K_{i}^{a}dx^{i}=e^{aj}K_{ij}dx^{i}$ is the extrinsic curvature
1-form and the antisymmetrization of indices in $e^{i[a}\partial
_{r}e_{i}^{b]}$ includes the factor $\frac{1}{2}$. The Lorentz connection
corresponds to the spacetime metric, $\omega ^{ab}(g)$, on the l.h.s. of the
equality, and to the boundary metric, $\omega ^{ab}(h)$, on the r.h.s.. The
induced metric $h_{ij}$ and its inverse $h^{ij}$ raise and lower the
boundary world indices, whereas the boundary vielbein $e_{i}^{a}$ and its
inverse $e_{a}^{i}$ projects the world indices $i,j,..$ to the Lorentz ones $%
a,b,..$, and vice versa.

With this notation and using $\epsilon _{1ab}=-\epsilon _{ab}$, the boundary
term is
\begin{eqnarray}
-\frac{k}{8\pi }\int\limits_{\partial \mathcal{M}}\epsilon _{ABC}\,\omega
^{AB}\wedge e^{C} &=&\frac{k}{8\pi }\int\limits_{\partial \mathcal{M}%
}\epsilon _{ab}\,\left( 2\omega ^{1a}\wedge e^{b}+\omega ^{ab}\wedge
e^{1}\right)  \notag \\
&=&\frac{k}{4\pi }\int\limits_{\partial \mathcal{M}}d^{2}x\,\epsilon
^{ij}\epsilon _{ab}\,K_{i}^{a}e_{j}^{b}=\frac{1}{2}\,B_{GH}\,.
\end{eqnarray}%
Note that $e^{1}=0$ and $\omega ^{ab}(g)=\omega ^{ab}(h)$\ on the boundary.
The Gibbons-Hawking boundary term reads
\begin{equation}
B_{GH}=-\frac{1}{8\pi G}\int\limits_{\partial \mathcal{M}}d^{2}x\sqrt{-h}%
\,K\,,  \label{GH}
\end{equation}%
\bigskip with $K=h^{ij}K_{ij}$ being the trace of the extrinsic curvature.

This calculation shows that the boundary term, which arises naturally in
Chern-Simons Poincar\'{e} gravity, equals one-half of the standard
Gibbons-Hawking boundary term, and we will use it as our boundary piece in
the gravitational action. In AdS gravity, this anomalous Gibbons-Hawking
boundary term \cite{Banados:1998ys} has been shown to result in a finite
action principle and proper values of the Noether charges \cite%
{Miskovic:2006tm,Banados:1994tn}.

The usual Gibbons-Hawking term provides a well-defined action principle for
the Dirichlet boundary conditions on the induced metric. A change of
boundary term has consequence of the boundary conditions, as well. In the
next section we address this question in asymptotically flat space.

\section{Boundary conditions}

A suitable set of boundary conditions for the action (\ref%
{csactionintermediate}) is the one for which the variation of the action
vanishes when the equations of motion hold. The variation of the action (\ref%
{actionCS}),\ on-shell, gives rise to a surface term%
\begin{eqnarray}
\delta I_{CS} &=&\frac{k}{4\pi }\int\limits_{\partial \mathcal{M}%
}\left\langle \delta A\wedge A\right\rangle  \notag \\
&=&\frac{k}{8\pi }\int\limits_{\partial \mathcal{M}}\epsilon _{ABC}\,\left(
\delta e^{A}\wedge \omega ^{BC}-e^{A}\wedge \delta \omega ^{BC}\right) \,.
\end{eqnarray}%
In an equivalent form, in an adapted frame (\ref{e}) which implies (\ref{w}%
), we have%
\begin{equation}
\delta I_{CS}=\frac{k}{4\pi }\int\limits_{\partial \mathcal{M}}\epsilon
_{ab}\,\left( \delta e^{a}\wedge \omega ^{1b}-e^{a}\wedge \delta \omega
^{1b}\right) \,,  \label{deltaICS}
\end{equation}%
for $r=Const$, in terms of boundary quantities.

Let us analyze the fall-off conditions in the boundary metric for a
spacetime which behaves asymptotically as a spinning particle ($\mu =-\alpha
^{2}$ in Eq. (\ref{metricConical})). The boundary is parametrized by the
local coordinates $x^{i}=(t,\theta )$, such that the induced metric behaves
for large $r$ as \cite{Ashtekar:1993ds,Marolf:2006xj}
\begin{equation}
h_{ij}=%
\begin{bmatrix}
-1+\mathcal{O}(1/r) & \mathcal{O}(r^{-\frac{\beta }{2}-1}) \\
\mathcal{O}(r^{-\frac{\beta }{2}-1}) & r^{2-\beta }+\mathcal{O}(r^{1-\beta })%
\end{bmatrix}%
.  \label{h_falloff}
\end{equation}%
One possible choice for the boundary zweibein is%
\begin{eqnarray}
e^{0} &=&A\,dt\,,  \notag \\
e^{2} &=&\frac{C}{r^{2}}\,dt+r^{1-\frac{\beta }{2}}B\,d\theta \,,
\label{e0,e2}
\end{eqnarray}%
where the functions $A(r,x)$,$\,B(r,x)$ and $C(r,x)$ are regular in the
asymptotic region, such that their expansion is%
\begin{eqnarray}
A &=&1+\mathcal{O}(1/r)\,,  \notag  \label{ABCseries} \\
B &=&1+\mathcal{O}(1/r)\,,  \notag \\
C &=&\mathcal{O}(1)\,.
\end{eqnarray}%
In addition, the lapse function for large $r$ has the form $N=r^{-\frac{%
\beta }{2}}+\mathcal{O}(r^{-\frac{\beta +1}{2}})$. The components of
Levi-Civit\`{a} connection $\omega ^{AB}(e)$ are%
\begin{eqnarray}
\omega ^{10} &=&-\frac{r^{\frac{\beta }{2}}A^{\prime }}{A}\,e^{0}-\chi
\,e^{2}\,,  \notag \\
\omega ^{12} &=&\chi \,e^{0}-r^{\frac{\beta }{2}}\left( \frac{B^{\prime }}{B}%
+\frac{2-\beta }{2r}\right) \,e^{2}\,,
\end{eqnarray}%
where the prime denotes radial derivative and we have defined the function
\begin{equation}
\chi =\frac{r^{\frac{\beta }{2}-2}}{2A}\left( \frac{CB^{\prime }-BC^{\prime }%
}{B}+\frac{\left( 6-\beta \right) C}{2r}\right) \,.
\end{equation}%
Asymptotically, the above function behaves as $\chi =\mathcal{O}(r^{\frac{%
\beta }{2}-3})$, what implies that $\omega ^{1a}$ behaves as
\begin{eqnarray}
\omega ^{10} &=&\mathcal{O}(r^{\frac{\beta }{2}-2})\,,  \notag \\
\omega ^{12} &=&-\frac{2-\beta }{2}\,Bd\theta +\mathcal{O}(1/r)\,.
\label{our w}
\end{eqnarray}%
On the other hand, the asymptotic form of the boundary frame is%
\begin{eqnarray}
e^{0} &=&A\,dt\,,  \notag \\
e^{2} &=&r^{1-\frac{\beta }{2}}B\,d\theta +\mathcal{O}(1/r^{2})\,,
\label{our e}
\end{eqnarray}%
what yields a finite variation of the action,%
\begin{equation}
\delta I_{CS}=-\frac{k}{4\pi }\frac{2-\beta }{2}\int d^{2}x\,\left( A\delta
B-B\delta A\right) \,.
\end{equation}
The action principle is satisfied if $A+\gamma B$ (with $\gamma =Const.$) vanishes on the boundary, because then $A\delta B-B\delta A=0$ on
$\partial \mathcal{M}.$ This condition is fulfilled since, from eq.(\ref{ABCseries}), $A=B$ up to the $\mathcal{O}(1/r)$ terms.

\section{Noether charge}

Let $L(\phi )$ be a Lagrangian 3-form describing a configuration of fields $%
\phi $, whose variation is $\delta L=\frac{\delta L}{\delta \phi }\,\delta
\phi +d\Theta (\phi ,\partial \phi ,\delta \phi )$, and $\xi =\xi ^{\mu
}\partial _{\mu }$ a set of asymptotic Killing vectors. The Noether current
corresponding to a diffeomorphism generated by the vector field $\xi ^{\mu
}(x)$ can be written in general as \cite{Iyer:1994ys}
\begin{equation}
\ast J=-\Theta -i_{\xi }L\,,  \label{WaldCurrent}
\end{equation}%
where $\ast J=\frac{1}{2}\sqrt{-g}\,\epsilon _{\mu \nu \lambda }J^{\mu
}dx^{\nu }\wedge dx^{\lambda }$ is the Hodge dual of the current. For the
Chern-Simons action (\ref{actionCS}), the above procedure for the connection
obeying the Chern-Simons equation of motion $F=dA+A\wedge A=0$ yields
\begin{equation}
\ast J=\frac{k}{4\pi }\,d\left\langle Ai_{\xi }A\right\rangle \,.
\label{CScurrent}
\end{equation}%
The above formula is a consequence of the fact that the diffeomorphisms
$\delta x^{\mu }=\xi ^{\mu }(x)$\ act on the fields as Lie derivatives, which
satisfy the differential geometry identity $\pounds _{\xi }=i_{\xi
}d+di_{\xi }$, where $i_{\xi }$\ is the contraction operator and $d=dx^{\mu
}\partial _{\mu }$\ is the exterior derivative. Thus, the Lie derivative
acts on the 3-form Lagrangian $L$\ as a total derivative $\pounds _{\xi
}L=d(i_{\xi }L)$. In consequence, invariance of the action $I[\phi ]=\int
L(\phi )$\ under general coordinate transformation implies the conservation
law $d\left. ^{\ast }J\right. =0$. For a given system, the Noether
current can always be written globally as $^{\ast }J=dQ[\xi ]$, as discussed in Ref.\cite{Barnich:2001jy},
such that  one can obtain the
Noether charge as a surface integral on the spacelike boundary $\partial
\Sigma $.

The charge is then expressed as an integral over an appropriate asymptotics,
\begin{equation}
Q[\xi ]=\frac{k}{4\pi }\int\limits_{\partial \Sigma }\left\langle Ai_{\xi
}A\right\rangle \,.  \label{CScharge}
\end{equation}

It is worthwhile noticing that general coordinate transformations with
parameter $\xi $ become algebraically equal, on-shell, to the Poincar\'{e}
gauge transformations upon field-dependent redefinitions of gauge
parameters: $\lambda ^{AB}=\xi ^{\nu }\omega _{\nu }^{AB}$ and $\lambda
^{A}=\xi ^{\nu }e_{\nu }^{A}$ for Lorentz rotations and translations,
respectively. Dependence of $\lambda $ on the gauge fields makes the
calculation of the conserved charges associated to Poincar\'{e}
transformations more complicated. A realization of off shell equivalence
between the two sets of local transformations involve trivial symmetries
\cite{Banerjee:2011cu}, which enables one to construct the charge (\ref%
{CScharge}) starting directly from $\mathfrak{iso(2,1)}$.

Indeed, a general coordinate transformation acting on the gauge connection is given
by the identity
\begin{equation}
\pounds_{\xi} A = D i_{\xi } A + i_{\xi } F\,,
\end{equation}
what makes evident that the gauge transformation is on-shell equivalent to a diffeomorphic transformation.
Therefore, the charges are the same. This is no longer true in higher-dimensional Chern-Simons theories \cite{Banados:1996yj}.

In the next section, we employ the equivalence between Chern-Simons theory
and gravity in $2+1$ dimensions to calculate the mass of the solutions in
asymptotically flat gravity.

\subsection{Conical singularity}

Let us study the conical singularity in the spinless case. We recall that
the metric is given by Eq.(\ref{metricConical}) with $\alpha >0$, where the
angular variable $\theta $ takes values $0\leq \theta \leq 2\pi $. The
angular deficit, $1-\alpha $, is related to a mass sitting at the
singularity through $\alpha =1-4Gm$ and the Minkowski vacuum corresponds to $%
\beta =0$ when the metric becomes identically flat with a full angular range
of $2\pi $, as discussed in (\ref{betaRange}).

We stress that the coordinate $r$ used in the metric (\ref{metricConical})
is not the usual radial distance from the center because the perimeter at $r$
is not $2\pi \left( 1-\alpha \right) r$. To get a locally flat metric (\ref%
{metricConicalAdeficit}) with the angular deficit, we have to change the
coordinates as (\ref{coordTransfconical}). On the other hand, the ADM form
of the metric with $N=\alpha ^{2}$ and $N_{\theta }=0$ is realized in the
ADM coordinates $\left( t^{\prime },r^{\prime },\theta \right) =(t/\alpha
,\alpha \rho ,\theta )$.

In a first-order description of the metric (\ref{metricConical}), we choose
the triad frame fields
\begin{equation}
e^{0}=dt\,,\qquad e^{1}=r^{-\frac{\beta }{2}}\,dr,\qquad e^{2}=r^{1-\frac{%
\beta }{2}}\,d\theta  \label{triadConic}
\end{equation}%
which, remembering that we have a torsionless and thus Riemannian manifold,
fixes the spin-connection through the triad postulate as,
\begin{equation}
\omega ^{12}=\frac{\beta -2}{2}\,d\theta .  \label{spinconnConic}
\end{equation}%
We now employ the CS formulation of $2+1$ gravity. Using the expression for
Noether charge corresponding to diffeomorphisms (\ref{CScharge}), mass is
given as the charge corresponding to the time translation Killing vector
field $\xi =\partial _{t}$,
\begin{equation}
Q[\partial _{t}]=\frac{k}{4\pi }\int\limits_{\partial \Sigma }\left\langle
\left( \frac{1}{2}\,\omega ^{AB}J_{AB}+e^{A}P_{A}\right) \left( \frac{1}{2}%
\,i_{\xi }\omega ^{AB}J_{AB}+i_{\xi }e^{A}P_{A}\right) \right\rangle .
\end{equation}%
Upon using the Poincar\'{e} algebra and the adopted trace $\left\langle
J_{AB}P_{C}\right\rangle =0$, we finally get
\begin{align}
Q[\partial _{t}]& =\frac{k}{4\pi }\int\limits_{0}^{2\pi }\frac{1}{2}%
\,\epsilon _{ABC}\left( \omega _{\theta }^{AB}e_{\ t}^{C}+\omega
_{t}^{AB}e_{\ \theta }^{C}\right) \,d\theta  \notag  \label{chargeConic} \\
& =\frac{k\,(\beta -2)}{4}\,.
\end{align}%
Thus the energy of the vacuum $(\beta =0)$ comes out to be
\begin{equation}
E_{0}=-\frac{k}{2}=-\frac{1}{8G}\,.  \label{QvacConic}
\end{equation}

\subsection{Cosmological asymptotically flat metric \label{Cosmo}}

In the previous section, we computed the vacuum energy as the Noether charge
for the conical singularity. Let us confirm that the vacuum energy does not
depend on the choice of the solution. Then, we consider the cosmological
asymptotically flat metric \cite{Barnich:2012xq} which lies in a different
sector of parameter space, $\mu =\alpha ^{2}$ and $j\neq 0$,
\begin{equation}
ds^{2}=-f^{2}dt^{2}+\frac{dr^{2}}{f^{2}}+r^{2}(d\theta +N_{\theta }\,dt)^{2}.
\label{metricBarnich}
\end{equation}%
Here $f^{2}(r)=-\mu +\frac{j^{2}}{r^{2}}$ and $N_{\theta }(r)=\frac{j}{r^{2}}
$.

To calculate the Noether charges, we follow a similar approach as outlined
in the previous section. The triad fields are chosen as
\begin{equation}
e^{0}=f\,dt,\quad e^{1}=\frac{1}{f}\,dr,\quad e^{2}=rN_{\theta
}\,dt+r\,d\theta ,  \label{triadBarnich}
\end{equation}%
which results in the torsionless spin connection
\begin{equation}
\omega ^{01}=-\frac{1}{2}\,r^{2}N_{\theta }^{\prime }\,d\theta \,,\quad
\omega ^{02}=-\frac{rN_{\theta }^{\prime }}{2f}\,dr,\quad \omega
^{12}=-f\,d\theta \,.  \label{spinconnBarnich}
\end{equation}%
Using (\ref{CScharge}), this gives corresponding to the killing vector
corresponding to time translations $\xi ^{t}=\partial _{t}$ a mass
\begin{equation}
Q[\partial _{t}]=4k\,GM\,.  \label{chargeBarnich}
\end{equation}%
The vacuum here is characterized by $J=0$ and$\ M=-\frac{1}{8G}$, because
then the metric becomes Minkowski. This results in the vacuum energy
\begin{equation}
E_{0}=-\frac{k}{2}=-\frac{1}{8G}\,,
\end{equation}%
what matches the result (\ref{QvacConic}).

To calculate the angular momentum, we just have to use the corresponding
angular Killing vector $\xi =\partial _{\theta }$ in (\ref{CScharge}),
\begin{align}
Q[\partial _{\theta }]& =\frac{k}{4\pi }\int\limits_{0}^{2\pi }\frac{1}{2}
\,\epsilon _{ABC}\left( \omega _{\theta }^{AB}e_{\ \theta }^{C}+\omega
_{\theta }^{AB}e_{\ \theta }^{C}\right) \,d\theta  \notag \\
& =4kGJ.
\end{align}%
Remembering that $k=\frac{1}{4G}$ leads to%
\begin{equation}
Q[\partial _{t}]=M\,,\qquad Q[\partial _{\theta }]=J\,,
\end{equation}%
as expected. We confirmed that the Noether charge formula (\ref{CScharge})
gives the correct values for the mass, $M$, and the angular momentum, $J$,
of the black hole and the vacuum energy, $E_{0}$.

At this point, we emphasize that it is the Chern-Simons
form of the action which leads to the correct answer for the charges in both flat and AdS cases,
what leaves no ambiguity in the choice of possible boundary terms.

\section{Discussion and Conclusions}

An inequivalent set of boundary conditions which accounts for conical
defects \cite{Deser:1983tn} and flat cosmologies \cite{Barnich:2012xq}
(discussed in Section \ref{Cosmo})\ in Euclidean sector with the line element%
\begin{equation}
ds^{2}=h_{\tau \tau }(\varphi )\,d\tau ^{2}+h_{rr}(\varphi )\,d\rho
^{2}+\rho ^{2}d\varphi ^{2}
\end{equation}%
is given by%
\begin{eqnarray}
\delta g_{\varphi \varphi } &=&\mathcal{O}(\rho )\,,\qquad \delta g_{\varphi
\tau }=\mathcal{O}(1)\,,\qquad \delta g_{\tau \tau }=\delta g_{\rho \rho }=%
\mathcal{O}(1)\,,  \notag \\
\delta g_{\tau \varphi } &=&\mathcal{O}(1)\,,\qquad \delta g_{\rho \tau }=%
\mathcal{O}(1/\rho )\,,\qquad \delta (g_{\rho \rho }g_{\tau \tau })=\mathcal{%
O}(1/\rho )\,.
\end{eqnarray}%
They are a particular case of the boundary conditions which are suitable to
treat asymptotically flat Einstein gravity \cite{Barnich:2006av} and
realizes Chiral Gravity in flat space\ \cite{Bagchi:2012yk}. In Ref.\cite%
{Detournay:2014fva} it was shown that the only way to have well-defined
action principle with this set of boundary conditions is to supplement the
action with a half of the Gibbons-Hawking term. From our point of view, this
choice is quite natural, as it is dictated by the Chern-Simons formulation
for $iso(2,1)$, that is, Eq.(\ref{csactionintermediate}). Therefore, the
conserved quantities constructed in the previous section can accommodate a
large class of solutions of flat gravity in three dimensions.

It is worthwhile noticing that these boundary conditions are suitable to
study 3D asymptotically flat Einstein gravity at null infinity, where the
asymptotic symmetries are described by the Bondi-Metzner-Sachs (BMS) group.
In general, BMS boundary conditions have a wave as a solution, and are
written in terms of the BMS coordinates that include retarded time, radius
and angle. A BMS gauge allows to treat the flat case as the limit \cite%
{Barnich:2012aw,Fareghbal:2013ifa} of the AdS case \cite{Coussaert:1995zp},
which is particularly useful to realize Flat/CFT\ correspondence \cite%
{Barnich:2010eb,Fareghbal:2014qga}. Furthermore, a 2D dual theory at null
infinity can be constructed starting from the CS formulation of 3D gravity
\cite{Barnich:2013yka}.

The construction presented here is inspired by, but differs from, the one
corresponding to Chern-Simons for AdS group. In three dimensions, a single
copy of Chern-Simons for $SO(2,2)$\ group gives rise to Einstein-Hilbert
action with negative cosmological constant \cite{Deser:1983nh} plus half of
the Gibbons-Hawking term \cite{Banados:1998ys}. It was shown in Ref.\cite%
{Miskovic:2006tm} that this boundary term renders the variation of the
action, at the same time, well defined and finite. The surface term in the
variation of the action adopts the same form as in Eq.(\ref{deltaICS}). At
first glance, it looks like one needs to impose a Neumann boundary condition
for the metric (i.e., fixing $K_{ij}$) for the action to be stationary \cite%
{Krishnan}. A posteriori, one can see that adding half of the
Gibbons-Hawking term is compatible with keeping a conformal structure at the
boundary, instead of the full boundary metric $h_{ij}$. In particular, this
can accommodate a holographic interpretation of the theory \cite%
{Miskovic:2006tm}. Indeed, the behavior of the fields in asymptotically AdS
gravity is such that the extrinsic curvature is proportional to the boundary
metric at leading order in the expansion. This accident happens only in the
AdS gravity: the absence of a conformal data in the boundary metric in
asymptotically flat gravity prevents a direct definition of holographic
quantities in this case.

\section*{Acknowledgments}

The authors would like to thank Glenn Barnich, Stephane Detournay and Hern%
\'{a}n Gonz\'{a}lez for useful comments. This work was supported by the
Chilean FONDECYT Grants N${{}^\circ}$3160139 and N${{}^\circ}$1131075,
VRIEA-PUCV grants N${{}^\circ}$039.345/2016 (O.M.), N${{}^\circ}$37.0/2015
(D.R.), UNAB Grant DI-1336-16/RG and CONICYT Grant DPI 20140115 (R.O.).

\end{document}